\documentclass[conference]{IEEEtran}
\IEEEoverridecommandlockouts

\usepackage{cite}
\usepackage{amsmath,amssymb,amsfonts}
\usepackage{algorithmic}
\usepackage{graphicx}
\usepackage{textcomp}
\usepackage{xcolor}
\usepackage{multirow}
\usepackage{amsfonts} 
\usepackage{amssymb} 
\usepackage{enumitem}
\usepackage{booktabs}
\usepackage{url}
\def\BibTeX{{\rm B\kern-.05em{\sc i\kern-.025em b}\kern-.08em
    T\kern-.1667em\lower.7ex\hbox{E}\kern-.125emX}}
\begin{document}

\title{Multimodal Contrastive Representation Learning Based Voice Conversion for Electrolaryngeal Speech Enhancement}

\newcommand{\equalcontrib}{\textsuperscript{*}}

\author{
\begin{minipage}{0.95\textwidth}
\centering

Ding Ma\textsuperscript{1,*},
Jinyi Mi\textsuperscript{1,*},
Jiajun He\textsuperscript{1,2},
Kazuhiro Kobayashi\textsuperscript{1,3},
Naoki Nishio\textsuperscript{4},
and Tomoki Toda\textsuperscript{5}

\textsuperscript{1}Graduate School of Informatics,
Nagoya University, Nagoya, Japan\\
\textsuperscript{2}Alibaba Token Foundry, Alibaba Group, Hangzhou, China\\
\textsuperscript{3}Headquarters, TARVO, Inc., Nagoya, Japan\\
\textsuperscript{4}Department of Otorhinolaryngology,
Graduate School of Medicine, Nagoya University, Nagoya, Japan\\
\textsuperscript{5}Information Technology Center,
Nagoya University, Nagoya, Japan\\

E-mail: ding.ma@g.sp.m.is.nagoya-u.ac.jp
\end{minipage}
\vspace{-0.5em}
}

\maketitle

\begingroup
\renewcommand{\thefootnote}{\fnsymbol{footnote}}
\footnotetext[1]{These authors contributed equally to this work.}
\endgroup

\begin{abstract}
    Electrolaryngeal (EL) speech suffers from severe acoustic mismatches with normal speech, resulting in degraded naturalness and intelligibility. These mismatches also make sequence-to-sequence voice conversion struggle to learn robust intermediate representations, hindering EL-speech-to-normal-speech conversion (EL2SP) performance. To address this, we propose a novel multimodal contrastive representation learning framework to 1) achieve richer multimodal-representation learning, and 2) facilitate the transferability of multimodal representations to speech-only EL2SP. First, EL speech and text representations are fused to learn multimodal representations for target speech generation, during which cross-modal contrastive learning is employed to encourage a semantically consistent speech--text representation space. Next, the fused representations are treated as a fixed target to guide speech-only representation learning through cross-stage contrastive learning, enabling multimodal knowledge to be transferred to the EL2SP task while preserving architectural and inference simplicity. Experimental results demonstrate that the proposed method consistently outperforms both the speech-only and multimodal-based baselines.
\end{abstract}

\vspace{-8pt}


\vspace{-6pt}
\section{Introduction}
Laryngectomy is a surgical treatment for laryngeal cancer~\cite{tang2015voice}, involving the removal of vocal folds. To speak without using vocal-fold vibration, laryngectomees commonly rely on portable electrolarynx as an external artificial excitation source. Its simulated vibration is transmitted into vocal tract to produce electrolaryngeal (EL) speech~\cite{singer1980endoscopic,hashiba2001industrialization,williams1985differences}. Nevertheless, EL speech is far from normal speech. The artificial excitation cannot reproduce dynamic characteristics of vocal-fold vibration, including rich fundamental frequency (F0) variation~\cite{ma1999improvement}, voiced/unvoiced cues~\cite{stanislav2017recognition}, and prosodic expressiveness. Consequently, EL speech sounds unnatural and has reduced phonetic clarity~\cite{violeta2024pretraining}. Meanwhile, high-energy artificial excitation inevitably radiates outside as noise, further degrading sound quality and intelligibility~\cite{kobayashi2018electrolaryngeal}. Furthermore, the compensatory speaking styles for these inherent limitations, such as slower articulation, would lead to abnormal speech duration and temporal structure. These factors result in substantial mismatches between EL and normal speech, causing communication burdens for both EL speakers and listeners.

Voice conversion (VC)~\cite{toda2007voice,stylianou1998continuous} originally aimed to convert speaker identity of speech from one speaker to another without changing linguistic content. VC can be used for EL-speech-to-normal-speech conversion (EL2SP). Sequence-to-sequence (seq2seq) modeling~\cite{sutskever2014sequence} is more suitable for EL2SP than conventional methods using strict frame-wise alignment~\cite{doi2013alaryngeal,tanaka2014hybrid,nakamura2012speaking,qian2021mandarin}. Seq2seq VC performs sequence-level, implicit mapping. It encodes linguistically intermediate representations from input sequence and then decodes them into target speech features. Thus, seq2seq VC can learn complex mappings and adjust the duration of converted speech to better match that of normal speech, which are essential for EL2SP.

To enhance seq2seq VC with limited EL2SP data in practice, Ma et al.~\cite{ma2025pretraining} introduced a pretraining and two-stage fine-tuning strategy with low-quality synthetic data, achieving then state-of-the-art performance. Nevertheless, this method faces performance bottlenecks due to large EL--normal mismatches, which introduce accumulated errors and unstable mappings.

\begin{figure*}[t]
\begin{center}
\includegraphics[width=0.85\linewidth]{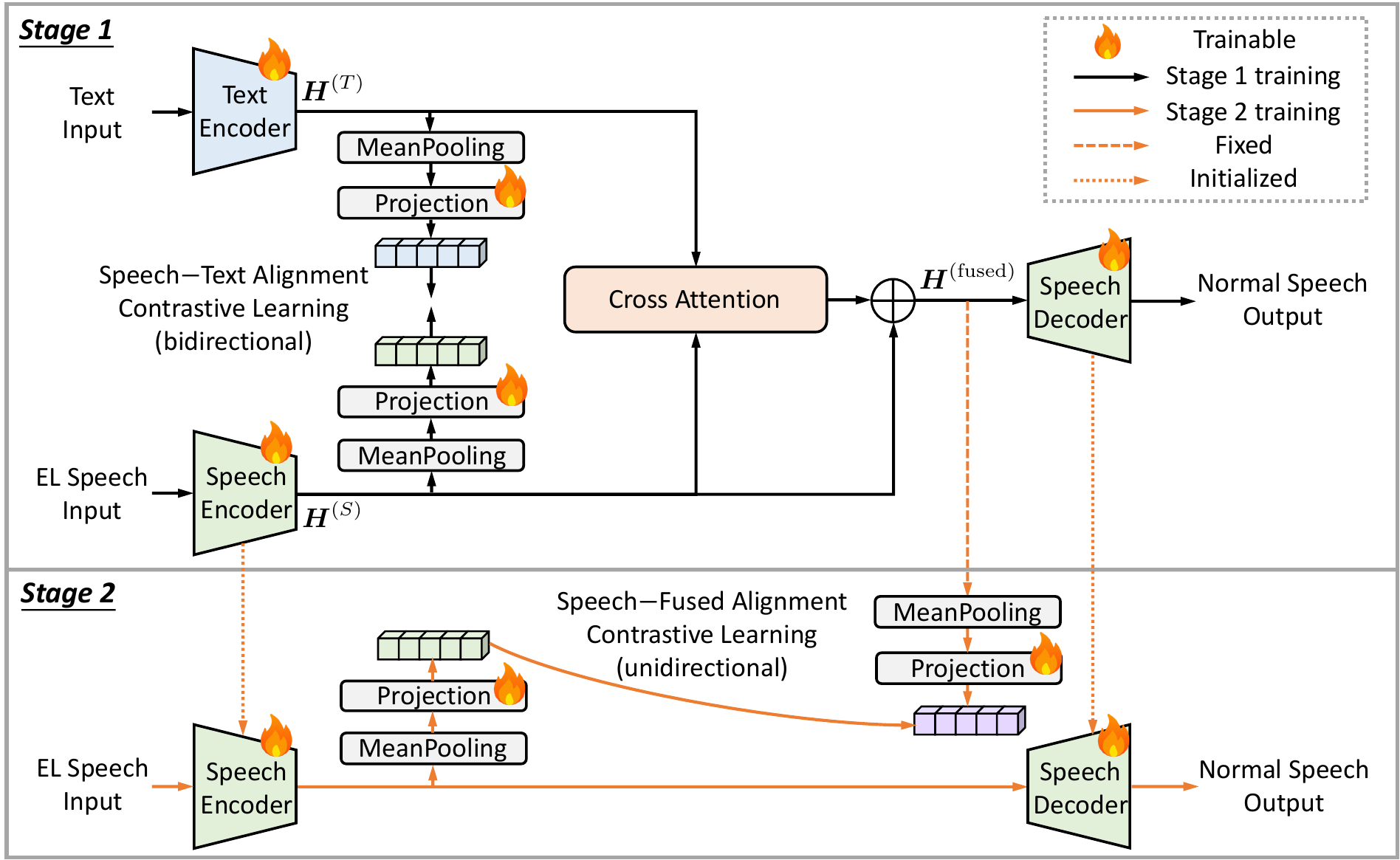}
\end{center}
\vspace{-15pt}
\caption{Overview of the proposed two-stage multimodal contrastive representation learning VC framework, MCRL-VC.}
\label{fig_framework}
\vspace{-15pt}
\end{figure*}

Incorporating text information provides a promising direction to improve the current approach. A direct approach is multimodal fusion, where additional modalities like text are explicitly integrated with speech representations. In VC tasks, Niu et al.~\cite{niu2024hybridvc} combined text and audio prompts for style VC, while Li et al.~\cite{li2023alignsts} introduced similar rhythm cues to enhance speech-to-singing conversion. Multimodal fusion strategies have also been investigated in broader speech processing tasks, such as speech emotion recognition~\cite{li2025wavfusion}, speech enhancement~\cite{10994671}, and automatic speech recognition~\cite{cappellazzo2025large}. While these studies demonstrate the effectiveness of integrating auxiliary modalities, they usually increase model complexity and data requirements, and often rely on additional modalities or prompts at run time. Such settings are not well aligned with practical EL2SP scenarios.

Motivated by these observations, this study exploits the text modality to enrich representation learning while preserving a simple architecture and inference pipeline. It focuses on two challenges. (1) Multimodal fusion: speech and text representations are heterogeneous, and EL speech contains degraded acoustic cues; thus, basic fusion is expected to be insufficient to capture fine-grained cross-modal correspondences. (2) Knowledge transfer: in EL2SP, multimodal knowledge should be transferred to a unimodal speech model for text-free inference.

We propose a novel two-stage \textbf{M}ultimodal \textbf{C}ontrastive \textbf{R}epresentation \textbf{L}earning \textbf{VC} (MCRL-VC) framework. First, paired speech and text representations are aligned into a shared semantic space, facilitating the multimodal fusion and mapping towards speech decoder. The learned fused representations are then effectively transferred to final EL2SP under contrastive learning guidance. We build different systems and conduct a comparative evaluation. 

The main contributions are as follows:
\begin{itemize}

\item We propose a two-stage multimodal contrastive representation learning framework for EL2SP. It enhances speech--text representation fusion and the transfer of multimodal representations.

\item Experimental results shows that the proposed method outperforms both a speech-only method~\cite{ma2025pretraining} and a multimodal method with the same modalities~\cite{11254511}.

\item We demonstrate the effectiveness of each stage in the proposed method.

\end{itemize}

\vspace{-5pt}
\section{Proposed Method}
\vspace{-5pt}
\subsection{Framework Overview}
In this work, we extend the concept of contrastive learning into a two-stage representation-guided learning framework, designed to improve the EL2SP task under a speech-only inference setting. Our method, MCRL-VC, first exploits the complementary information between the speech and text modalities, through \textit{cross-modal contrastive learning} to get more effective representation fusion. Then, using \textit{cross-stage contrastive learning}, the fused knowledge is transferred back to a standalone speech encoder.

Following the above concept, as illustrated in Fig.~\ref{fig_framework}, \textit{Stage~1} focuses on aligning acoustic information from EL speech with linguistic information from text, and integrating them through cross-attention mechanism (Section~\ref{method_stage1}). \textit{Stage~2} aims to transfer the integrated knowledge learned in \textit{Stage~1} into a speech encoder. During this process, the integrated representations learned in \textit{Stage~1} serves as a teacher, guiding the speech encoder and enabling speech-only reconstruction (Section~\ref{method_stage2}).

\vspace{-4pt}
\subsection{Stage 1: Cross-modal Text-guided Representation Learning}
\label{method_stage1}

The first stage involves a joint network consisting of a speech encoder, a text encoder, a speech decoder, and a cross-attention layer. Let $\{(X^{(S)}_i,X^{(T)}_i)\}_{i=1}^{N}$ denote paired EL speech and text samples, where \(N\) is the batch size, \(S\) and \(T\) indicate the speech and text modalities, respectively. The speech and text encoders extract modality-specific representations as
\begin{align}
    \boldsymbol{H}^{(S)} = \mathrm{SpeechEnc}(\boldsymbol{X}^{(S)}) 
    \in \mathbb{R}^{N \times F \times D}, \\
    \boldsymbol{H}^{(T)} = \mathrm{TextEnc}(\boldsymbol{X}^{(T)}) 
    \in \mathbb{R}^{N \times L \times D},
\end{align}
where $F$ and $L$ denote the number of speech frames and text-sequence length, respectively, and $D$ is the hidden dimension.

To facilitate effective cross-modal fusion, we utilize the cross-modal contrastive learning to align the representations from EL speech and text. Specifically, the contrastive objective encourages paired speech and text representations to be close in the latent space, while pushing unmatched pairs apart. We separately process the speech representation \(\boldsymbol{H}^{(S)}\) and the text representation \(\boldsymbol{H}^{(T)}\) by applying mean pooling along the temporal dimension to obtain utterance-level representations. The pooled representations are then fed into modality-specific projection blocks, which map them into a shared contrastive space with the same dimensionality. This enables the computation of a cross-modal similarity matrix between~\(\boldsymbol{H}^{(S)}\) and~\(\boldsymbol{H}^{(T)}\). These processes are formulated as
\begin{align}
  {\boldsymbol{Z}^{(S)} = \mathrm{Proj}(\mathrm{MeanPooling}({\boldsymbol{H}^{(S)}))}} \in \mathbb{R}^{N \times D}, \\
  {\boldsymbol{Z}^{(T)} = \mathrm{Proj}(\mathrm{MeanPooling}({\boldsymbol{H}^{(T)}))}} \in \mathbb{R}^{N \times D},
\end{align}
where \(\mathrm{Proj}(\cdot)\) denotes trainable projection modules, each consisting of a linear layer followed by layer normalization.
The cross-modal similarity matrices for speech-to-text and text-to-speech matching are then computed as
\begin{align}
    \boldsymbol{M}^{(S \rightarrow T)}
    =
    \hat{\boldsymbol{Z}}^{(S)}
    (\hat{\boldsymbol{Z}}^{(T)})^{\top}/ \tau
    \in \mathbb{R}^{N \times N}, \\
    \boldsymbol{M}^{(T \rightarrow S)}
    =
    \hat{\boldsymbol{Z}}^{(T)}
    (\hat{\boldsymbol{Z}}^{(S)})^{\top}/ \tau
    \in \mathbb{R}^{N \times N}, 
\end{align}
where $\tau$ is a temperature parameter, $\hat{\boldsymbol{Z}}^{(S)}$ and $\hat{\boldsymbol{Z}}^{(T)}$ denote the L2-normalized \(\boldsymbol{Z}^{(S)}\) and \(\boldsymbol{Z}^{(T)}\), respectively. The speech-to-text and text-to-speech contrastive losses are based on a common InfoNCE-style objective \cite{ye2022cross} and are defined as
\begin{align}
    \mathcal{L}_{S \rightarrow T}
    &=
    -\frac{1}{N}
    \sum_{i=1}^{N}
    \log
    \frac{
    \exp(\boldsymbol{M}^{(S \rightarrow T)}_{i,i})
    }{
    \sum_{j=1}^{N}
    \exp(\boldsymbol{M}^{(S \rightarrow T)}_{i,j})
    }, \\
    \mathcal{L}_{T \rightarrow S}
    &=
    -\frac{1}{N}
    \sum_{i=1}^{N}
    \log
    \frac{
    \exp(\boldsymbol{M}^{(T \rightarrow S)}_{i,i})
    }{
    \sum_{j=1}^{N}
    \exp(\boldsymbol{M}^{(T \rightarrow S)}_{i,j})
    },
\end{align}
where $\boldsymbol{M}^{(S \rightarrow T)}_{i,j}$ measures the similarity between the $i$-th speech representation and the $j$-th text representation, while $\boldsymbol{M}^{(T \rightarrow S)}_{i,j}$ measures the similarity between the $i$-th text representation and the $j$-th speech representation. The final cross-modal contrastive (CMC) loss is obtained by averaging the two directional losses:
\begin{equation}
    \mathcal{L}_{\mathrm{CMC}}
    =
    \frac{1}{2}
    \left(
    \mathcal{L}_{S \rightarrow T}
    +
    \mathcal{L}_{T \rightarrow S}
    \right).
\end{equation}

Meanwhile, we utilize a multi-head cross-attention block to incorporate linguistic information from \(\boldsymbol{H}^{(T)}\) into the speech representation \(\boldsymbol{H}^{(S)}\), where $\boldsymbol{H}^{(S)}$ is used as the query, and $\boldsymbol{H}^{(T)}$ serves as the key and value. A residual connection is then applied to reintegrate the original speech representation $\boldsymbol{H}^{(S)}$ for information integrity. This process is formulated as
\begin{equation}
\begin{aligned}
    \boldsymbol{H}^{(\mathrm{fused})}
    &=
    \mathrm{CrossAttention}
    \big(
    \boldsymbol{H}^{(S)}, 
    \boldsymbol{H}^{(T)}, 
    \boldsymbol{H}^{(T)}
    \big)
    + \boldsymbol{H}^{(S)}  \\
    &\in \mathbb{R}^{N \times F \times D}.
\end{aligned}
\end{equation}

\vspace{-5pt}
\subsection{Stage 2: Cross-stage Multimodal Representation Transfer}
\label{method_stage2}

In the second stage, we remove the text encoder and construct a speech-only EL2SP model by initializing the speech encoder and decoder with their corresponding parameters trained in the first stage. To this end, the text-enhanced representation $\boldsymbol{H}^{(\mathrm{fused})}$ from the first stage is frozen and used as a representation-level teacher to construct the contrastive loss, thus guiding the speech-only EL2SP training.

Given the EL speech input $\boldsymbol{X}^{(S)}$, the speech encoder extracts the student speech representation as
\begin{equation}
    \boldsymbol{H}^{(\mathrm{stu})}
    =
    \mathrm{SpeechEnc}(\boldsymbol{X}^{(S)})
    \in \mathbb{R}^{N \times F \times D}.
\end{equation}

To perform cross-stage representation transfer, both student and teacher representations are mean-pooled along the temporal dimension and then mapped into a shared contrastive space through a trainable projection block:
\begin{align}
    \boldsymbol{Z}^{(\mathrm{stu})}
    &=
    \mathrm{Proj}(\mathrm{MeanPooling}
    (\boldsymbol{H}^{(\mathrm{stu})}))
    \in \mathbb{R}^{N \times D}, \\
    \boldsymbol{Z}^{(\mathrm{fused})}
    &=
    \mathrm{Proj}(\mathrm{MeanPooling}
    (\boldsymbol{H}^{(\mathrm{fused})}))
    \in \mathbb{R}^{N \times D}.
\end{align}
The cross-stage similarity matrix is computed as
\begin{equation}
    \boldsymbol{M}^{(\mathrm{stu} \rightarrow \mathrm{fused})}
    =
    \hat{\boldsymbol{Z}}^{(\mathrm{stu})}
    (\hat{\boldsymbol{Z}}^{(\mathrm{fused})})^{\top}/ \tau
    \in \mathbb{R}^{N \times N},
\end{equation}
where $\hat{\boldsymbol{Z}}^{(\mathrm{stu})}$ and $\hat{\boldsymbol{Z}}^{(\mathrm{fused})}$ denote the L2-normalized ${\boldsymbol{Z}}^{(\mathrm{stu})}$ and ${\boldsymbol{Z}}^{(\mathrm{fused})}$, respectively. Then, we apply a unidirectional cross-stage contrastive (CSC) loss from the student speech representation to the teacher representation:
\begin{equation}
    \mathcal{L}_{\mathrm{CSC}}
    =
    -\frac{1}{N}
    \sum_{i=1}^{N}
    \log
    \frac{
    \exp(
    \boldsymbol{M}^{(\mathrm{stu} \rightarrow \mathrm{fused})}_{i,i})
    }{
    \sum_{j=1}^{N}
    \exp(
    \boldsymbol{M}^{(\mathrm{stu} \rightarrow \mathrm{fused})}_{i,j})
    }.
\end{equation}
This CSC loss encourages the speech encoder in the second stage to approximate the fused representation learned in the first stage without requiring text input. Some related contrastive distillation strategies have been explored in \cite{Tian2020con,Afouras2020asr}.

\subsection{Training Loss}

The model is typically optimized using three types of losses: the L1 loss for target speech reconstruction, the weighted binary cross-entropy (BCE) loss for stop token prediction, and the guided attention (GA) loss for encouraging a more effective diagonal alignment between the input and output sequences.
In our framework, we further introduce the CMC loss and CSC loss in the first and second stages, respectively. The overall losses for the two stages are formulated as
\begin{align}
    \mathcal{L}_{\mathrm{Stage 1}}
    =
    \mathcal{L}_{\mathrm{L1}}
    +
    \mathcal{L}_{\mathrm{BCE}}
    +
    \mathcal{L}_{\mathrm{GA}}
    +
    \alpha\mathcal{L}_{\mathrm{CMC}}, \\
    \mathcal{L}_{\mathrm{Stage 2}}
    =
    \mathcal{L}_{\mathrm{L1}}
    +
    \mathcal{L}_{\mathrm{BCE}}
    +
    \mathcal{L}_{\mathrm{GA}}
    +
    \beta\mathcal{L}_{\mathrm{CSC}},
\end{align}
where $\alpha$ and $\beta$ control the contributions of the two contrastive objectives.

\vspace{-3pt}
\section{Experimental Settings}
\subsection{System Building}

We compare the proposed system, \textit{MCRL-VC}, with two baselines: a Speech-only VC (\textit{S-VC}), and a Multimodal-Representation-based VC (\textit{MR-VC}). We also construct a variant of \textit{MCRL-VC}, referred to as \textit{MCRL-VC-Rec}.

\begin{itemize}[label=\textbullet, leftmargin=\parindent, labelindent=\parindent]

\item \textit{S-VC} follows the VC pretraining and two-stage fine-tuning method as in~\cite{ma2025pretraining}. In VC pretraining, a pretrained TTS model serves as the source model: its decoder is transferred to VC, and a speech encoder is pretrained via fixed-decoder speech autoencoding~\cite{huang2019voice}. The first fine-tuning stage incorporates data augmentation with parallel synthetic data~(SD), together with the original speech data. The SD are generated by feeding an external text set into EL and normal TTS models, both initialized from the same source TTS model and fine-tuned separately. The second stage further refines model parameters using only the original speech data.

\item \textit{MR-VC} follows the speech--text representation learning method in~\cite{11254511}. In the first training stage, a joint network is initialized with the pretrained VC and normal TTS modules prepared in S-VC, and trained using the same SD and original speech data, as well as paired text. EL-speech and text representations are fused through cross-attention mechanism for normal speech generation. In the second stage, the text encoder is removed, and direct speech-only reconstruction training is performed using original speech data to obtain the final VC model.

\item \textit{MCRL-VC} follows the proposed method in Section~II. It uses the same training data settings, pretrained VC and normal TTS modules as MR-VC.

\item \textit{MCRL-VC-Rec} retains the first-stage contrastive learning of MCRL-VC, but follows the reconstruction training of MR-VC in the second stage without introducing the cross-stage contrastive learning.

\end{itemize}

\vspace{-10pt}
\subsection{Datasets}
Two types of Japanese datasets were used to construct the proposed systems.
\subsubsection{TTS Database} 
The JSUT corpus~\cite{sonobe2017jsut} was used for VC pretraining. It contains 7296 training utterances, totaling approximately 10 hours. In addition, the JSUT transcriptions and the SD generated using them both contributed to the first-stage training.

\subsubsection{EL2SP Datasets}
We used two small-scale EL2SP datasets, referred to as the Patient 1 and Patient 2 datasets.

\begin{itemize}[label=\textbullet, leftmargin=\parindent, labelindent=\parindent]
\item \textit{Patient 1 dataset} is a parallel corpus containing 300 EL and normal speech utterances, with a total duration of roughly 15 minutes. The normal and EL speech were recorded from a patient before and after total laryngectomy, respectively.

\item \textit{Patient 2 dataset} is a semi-parallel corpus containing 200 EL speech utterances of less than 10 minutes and 413 normal speech utterances of about 20 minutes. The EL speech was recorded from a patient who underwent total laryngectomy. Since the corresponding preoperative normal speech was unavailable, a healthy speaker provided the normal speech utterances under the same recording conditions. This dataset simulates a practical scenario where the patient's own normal speech cannot be obtained.
\end{itemize}

For both datasets, 20 parallel utterances were used the development set and 40 as the test set, while the remaining utterances were used for training. Specifically, for the first-stage fine-tuning of S-VC and the joint network training of MR-VC, MCRL-VC, and MCRL-VC-Rec, besides the original parallel pairs (240 for Patient 1 and 140 for Patient 2), we incorporated external parallel SD (7296). For semi-parallel Patient 2 dataset, since the normal speech contained additional utterances without real EL counterparts, corresponding EL SD were generated for them, enabling all normal speech data (353) to be used in these stages. In their final stages, only the original parallel data were reused for training.

\vspace{-4pt}
\subsection{Implementations}
The experimental systems were implemented using the ESPnet toolkit~\cite{watanabe2018espnet}. All speech signals were resampled at 24~kHz, and 80-dimensional mel-filterbank features were extracted with 2048 FFT points and a 300-point frame shift. The input text feature sequences for the text encoders consisted of Japanese phonemes and pause symbols. The encoders and decoders used six-layer Transformer blocks with four attention heads and a reduction factor of 3. The cross-attention block in the joint network also used four attention heads. The learning rate was set to 0.001, and the Lamb~\cite{you2020large} and Noam~\cite{NIPS2017_3f5ee243} optimizers were used for VC and TTS-related training, respectively. For MCRL-VC, we set $\alpha=0.1$ and $\beta=0.01$ in Eqs.~(16) and~(17), respectively. To reconstruct waveforms of the SD and EL2SP outputs, four speech-dependent Parallel WaveGAN vocoders~\cite{yamamoto2020parallel} were trained following the original configurations, separately for EL and normal speech in the Patients 1 and 2 datasets.

\vspace{-4pt}
\subsection{Evaluation Metrics} 

 We used four objective metrics: Mel-cepstral distortion (MCD) measures spectral distortion between converted and normal speech. Character error rate (CER, in \%) was calculated using an EL- and normal-speech adapted ASR engine~\cite{violeta2022investigating} to evaluate intelligibility. Speaker embedding cosine similarity (SECS) was measured between converted and reference speech embeddings extracted using a speaker verification tool from SpeechBrain~\cite{speechbrain}. Subjective speech quality assessment (SSQA) was used through a mean opinion score (MOS) predictor~\cite{11488533} to automatically estimate perceived naturalness on a five-point scale. For subjective evaluation, twelve utterances from each system were randomly selected, and 26 native Japanese listeners were recruited to assess the naturalness with MOS (NMOS). Audio samples are available online\footnote{\url{https://muginekoo.github.io/slt2026-demo/}}.

\begin{table*}[t]\small
\caption{Objective and subjective evaluation results on the comparable systems, original EL, and normal (NL) speech for Patient 1, with 95\% confidence intervals shown for NMOS.}\label{tab:results_patient1}
\vspace{-4pt}
\centering
\setlength{\tabcolsep}{8pt}
{\fontsize{9.5pt}{11pt}\selectfont
\begin{tabular}{lccccc}
\toprule[1pt]
{\textbf{Systems}} & {\textbf{MCD} ($\downarrow$)} & {\textbf{CER} ($\downarrow$)} & {\textbf{SECS} ($\uparrow$)} & {\textbf{SSQA} ($\uparrow$)} & {\textbf{NMOS} ($\uparrow$)} \\ \midrule
S-VC      & 7.42      & 37.0      & 0.587      & 3.80       & 2.64 $\pm$ 0.09 \\
MR-VC     & 7.35      & 35.5      & 0.630      & 3.83       & 3.04 $\pm$ 0.10 \\
MCRL-VC (proposed) & \textbf{7.20}      & \textbf{32.7}      & \textbf{0.647}      & 3.84       & \textbf{3.15} $\pm$ \textbf{0.09} \\ 
MCRL-VC-Rec (proposed) & 7.25      & 33.9      & 0.644      & \textbf{3.88}       & -- \\ \midrule
EL        & 13.28     & 41.8      & 0.014      & 2.12       & 1.08 $\pm$ 0.04 \\
NL (ground truth) & -- & 15.1      & --          & 4.48       & 4.62 $\pm$ 0.07 \\
\bottomrule[1pt]
\end{tabular}}
\vspace{-10pt}
\end{table*}

\begin{table*}[t]\small
\caption{Objective and subjective evaluation results on the comparable systems, original EL, and normal (NL) speech for Patient 2, with 95\% confidence intervals shown for NMOS.}
\label{tab:results_patient2}
\vspace{-4pt}
\centering
\setlength{\tabcolsep}{8pt}
{\fontsize{9.5pt}{11pt}\selectfont
\begin{tabular}{lccccc}
\toprule[1pt]
{\textbf{Systems}} & {\textbf{MCD} ($\downarrow$)} & {\textbf{CER} ($\downarrow$)} & {\textbf{SECS} ($\uparrow$)} & {\textbf{SSQA} ($\uparrow$)} & {\textbf{NMOS} ($\uparrow$)} \\ \midrule
S-VC      & 6.24      & 23.3      & 0.734      & 3.91       & 2.56 $\pm$ 0.09 \\
MR-VC     & 6.04      & 22.1      & 0.738      & 3.93       & 2.86 $\pm$ 0.09 \\
MCRL-VC (proposed) & \textbf{5.87}      & \textbf{20.1}      & \textbf{0.743}      & 3.95       & \textbf{3.28} $\pm$ \textbf{0.10} \\ 
MCRL-VC-Rec (proposed) & 5.96      & 21.6      & 0.739      & \textbf{3.99}       & -- \\ \midrule
EL        & 12.40     & 25.1      & 0.047      & 2.76       & 1.18 $\pm$ 0.05 \\
NL (ground truth) & -- & 10.9      & --          & 4.22       & 4.41 $\pm$ 0.09 \\
\bottomrule[1pt]
\end{tabular}}
\vspace{-14pt}
\end{table*}

\begin{figure*}[]
    \centering
    \begin{minipage}[t]{1.0\textwidth}
        \centering
        \includegraphics[width=\linewidth, trim=0 4mm 0 0, clip]{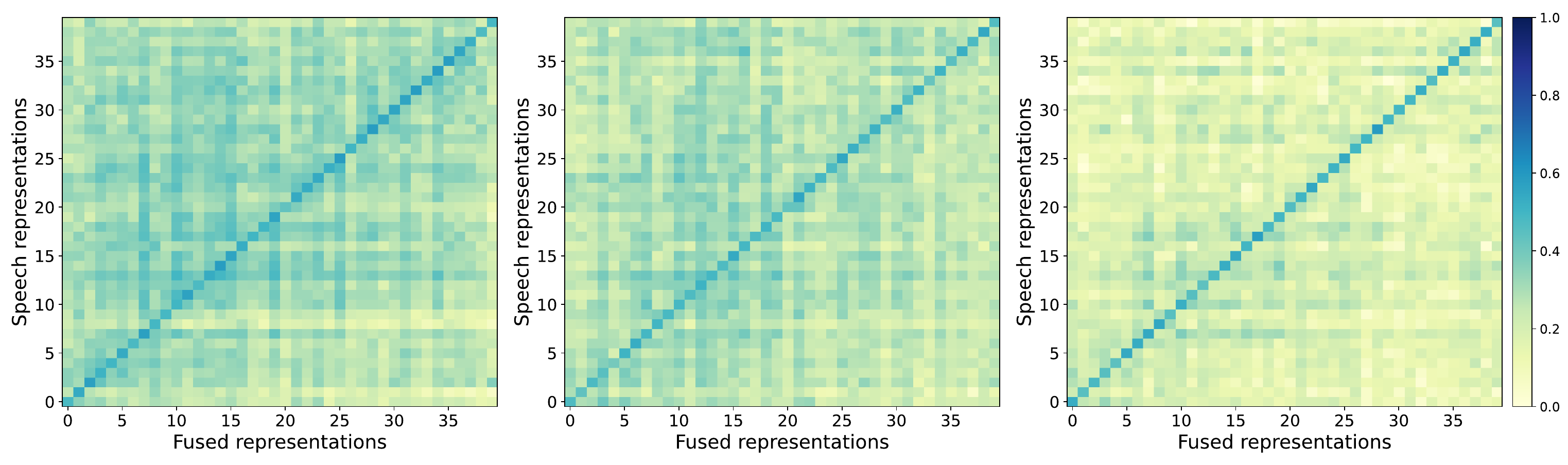}\\[-5pt]

        \noindent\makebox[\linewidth]{%
            \scriptsize
            \makebox[0.33\linewidth][c]{(a) MR-VC}%
            \hfill
            \makebox[0.33\linewidth][c]{(b) MCRL-VC-Rec}%
            \hfill
            \makebox[0.35\linewidth][c]{(c) MCRL-VC}%
        }\medskip
    \end{minipage}
    \vspace{-15pt}
    \caption{Heatmaps of cosine similarity matrices between the second-stage speech representations and the corresponding first-stage fused representations within each system: (a) MR-VC, (b) MCRL-VC-Rec, and (c) MCRL-VC, using the test set from the Patient 1 dataset.}
    \label{fig:heatmap}
    \vspace{-15pt}
\end{figure*}

\vspace{-8pt}
\section{Experimental Results}
\subsection{Comparison of MCRL-VC and Baselines}

Tables~\ref{tab:results_patient1} and~\ref{tab:results_patient2} summarize the objective and subjective results for Patients 1 and 2, respectively. For the comparison between the two baselines, MR-VC consistently outperforms S-VC across all objective and subjective evaluations, indicating that speech--text representation learning is more effective than relying on speech-only representation learning. 

Next, we compare MCRL-VC and MR-VC in objective results. MCRL-VC achieves better overall performance than MR-VC. Specifically, compared with MR-VC, MCRL-VC reduces MCD by 0.15 and CER by 2.8\% on Patient 1, and reduces MCD by 0.17 and CER by 2.0\% on Patient 2. A similar trend can be observed in the subjective results. Especially, on the Patient 2 dataset, MCRL-VC obtains a significantly higher NMOS than MR-VC, showing the closest naturalness to the ground truth speech. We expect that this improvement comes from two aspects. First, through cross-modal contrastive learning, MCRL-VC can capture more fine-grained cross-modal information during multimodal fusion than the basic fusion strategy used in MR-VC. Second, the cross-stage contrastive learning strategy provides more robust guidance for transferring multimodal knowledge to the final VC model, thereby improving the overall conversion performance.

\vspace{-5pt}
\subsection{Stage-wise Validation of Contrastive Learning}
\vspace{-2pt}

To further verify the above expectations, we analyze the contribution of the stage-wise design. As shown in Tables~\ref{tab:results_patient1} and~\ref{tab:results_patient2}, we first compare MCRL-VC-Rec with MR-VC. MCRL-VC-Rec clearly outperforms MR-VC. Since MCRL-VC-Rec additionally introduces cross-modal contrastive learning in the first stage, this result supports our expectation that this strategy can enhance the effectiveness of multimodal fusion. We then compare MCRL-VC-Rec with MCRL-VC. MCRL-VC achieves better performance than MCRL-VC-Rec on three out of the four evaluation metrics, further verifying the effectiveness of the cross-stage contrastive learning in the second stage. We also notice that MCRL-VC performs slightly worse than MCRL-VC-Rec in SSQA, showing a potential trade-off on speech quality.

\vspace{-2pt}
Furthermore, we visualized representation-level alignment for the three systems in Fig~\ref{fig:heatmap}, by computing cosine similarity matrices between mean-pooled, L2-normalized speech representations (second stage) and fused representations (first stage). Compared with MR-VC, MCRL-VC-Rec exhibits a clearer diagonal pattern and weaker off-diagonal similarities, suggesting that cross-modal contrastive learning produces fused representations that are more compatible with second-stage speech representations. This enables smoother transfer to speech-only reconstruction. MCRL-VC shows the strongest diagonal dominance and the lowest off-diagonal similarities among the three systems. This indicates that teacher guidance via cross-stage contrastive learning further encourages speech-only representations to align more closely with the fused representations. These trends are similar with those in Table~\ref{tab:results_patient1}, showing the effectiveness of such stage-wise design.

\vspace{-3pt}
\section{Conclusion}
\vspace{-2pt}

In the work, we proposed MCRL-VC, a novel two-stage multimodal contrastive representation learning VC framework, involving cross-modal and cross-stage contrastive learning, to realize enhanced EL2SP without requiring text input. Specifically, the proposed method addressed the two key challenges in EL2SP tasks: how to achieve more effective multimodal fusion and more reliable knowledge transfer. Experimental results showed that the proposed method outperformed both speech-only and multimodal-based VC baselines. Moreover, the stage-wise variant analysis further confirmed the contribution of each stage. Future work will focus on more practical real-world applications, including (1) incorporating additional modalities, and (2) developing low-latency EL2SP systems.
\vspace{-2pt}
\section*{Acknowledgment}
\vspace{-2pt}
This work was supported in part by the BRIDGE Program (R7-H05), implemented by the Cabinet Office,
Government of Japan, and in part by JSPS KAKENHI under Grant 25K02788, Japan.

The authors used generative AI tools (ChatGPT) for grammar checking and correction, and improvement of clarity and readability of the manuscript.

\vspace{-5pt}
\bibliographystyle{IEEEtran}
\bibliography{refs}

\end{document}